\documentclass[english]{interact}
\usepackage[T1]{fontenc}
\usepackage[latin9]{inputenc}
\usepackage{refstyle}
\usepackage{mathrsfs}
\usepackage{amsmath}
\usepackage{graphicx}

\makeatletter

\AtBeginDocument{\providecommand\figref[1]{\ref{fig:#1}}}
\AtBeginDocument{\providecommand\secref[1]{\ref{sec:#1}}}
\AtBeginDocument{\providecommand\subsecref[1]{\ref{subsec:#1}}}
\RS@ifundefined{subsecref}
  {\newref{subsec}{name = \RSsectxt}}
  {}
\RS@ifundefined{thmref}
  {\def\RSthmtxt{theorem~}\newref{thm}{name = \RSthmtxt}}
  {}
\RS@ifundefined{lemref}
  {\def\RSlemtxt{lemma~}\newref{lem}{name = \RSlemtxt}}
  {}

\usepackage{pgfplots}
\usepgfplotslibrary{groupplots}
\usepackage{mhchem}

\renewcommand{\figref}{\Figref}
\usepackage[numbers]{natbib}

\makeatother

\usepackage{babel}
\begin{document}

\title{Accelerating the Convergence of Higher-Order Coupled Cluster Methods
II: Coupled Cluster $\Lambda$ Equations and Dynamic Damping}

\author{\name{Devin A. Matthews$^\ast$\thanks{$^\ast$E-mail address: damatthews@smu.edu}}\affil{Southern Methodist University, Dallas, TX 75275}}
\maketitle
\begin{abstract}
The method of sub-iteration, which was previously applied to the higher-order
coupled cluster amplitude equations, is extended to the case of the
coupled cluster $\Lambda$ equations. The sub-iteration procedure
for the $\Lambda$ equations is found to be highly similar to that
for the amplitude equations, and to exhibit a similar improvement
in rate of convergence relative to extrapolation of all $\hat{T}$
or $\hat{\Lambda}$ amplitudes using DIIS. A method of dynamic damping
is also presented which is found to effectively recover rapid convergence
in the case of oscillatory behavior in the amplitude or $\Lambda$
equations. Together, these techniques allow for the convergence of
both the amplitude and $\Lambda$ equations necessary for the calculation
of analytic gradients and properties of higher-order coupled cluster
methods without the high memory or disk I/O cost of full DIIS extrapolation.
\end{abstract}

\section{Introduction}

Higher-order coupled cluster methods,\cite{hirataHighorderCoupledclusterCalculations2000,kallayHigherExcitationsCoupledcluster2001,kallayApproximateTreatmentHigher2005}
that is, methods beyond the coupled cluster singles and doubles level
(CCSD),\cite{ccsd-1,ccsd-2} are important tools for obtaining extremely
accurate theoretical predictions of thermochemical and kinetic parameters.\cite{thermo5,thermo1,thermo2,thermo4,thermo3,thermo7,thermo6}
Additionally, such methods provide reliable benchmark values for the
development of lower-cost approximations. The application of non-iterative
approximations to such methods, in particular the ``gold standard''
CCSD(T) method,\cite{raghavachariFifthorderPerturbationComparison1989a}
require only sufficient computational power in order to complete the
calculation. More elaborate approximations, however, and in particular
the ``full'' higher-order coupled cluster methods CCSDT\cite{ccsdt}
and CCSDTQ,\cite{oliphantCoupledClusterMethod1991,kucharskiRecursiveIntermediateFactorization1991,ccsdtq}
present additional obstacles in that actually converging the coupled
cluster equations becomes more difficult. Convergence of the CCSDT
and CCSDTQ equations using a simple ``Jacobi-like'' iteration is
essentially impossible, and even with extrapolation methods like RLE,\cite{rle-1,rle-2}
DIIS,\cite{diis-1,diis-2,diis-scf,diis-cc} and CROP,\cite{crop-1,crop-2}
the number of iterations required for satisfactory convergence of
these equations is almost always higher than for CCSD. More concerning
are the cases where, despite all attempts at extrapolation, the equations
will simply not converge.

In a previous publication,\cite{matthewsAcceleratingConvergenceHigherorder2015}
which we refer to as Paper I, we presented a method based on ``sub-iteration''
that a) avoids the costly extrapolation of the higher-order amplitudes
when using an extrapolation method such as DIIS, and b) provides for
more rapid convergence in the general case, even compared to full
DIIS. In this paper, we extend this method to the case of the coupled
cluster $\Lambda$ equations which are a necessary step in the calculation
of analytic gradients and properties.\cite{cc-gradient-1,cc-gradient-2,kallayAnalyticFirstDerivatives2003}
Calculations of such properties using higher-order coupled cluster
methods are becoming increasingly important, for example in the use
of composite schemes for geometry optimization\cite{morganGeometricEnergyDerivatives2018}
which can provide extremely accurate equilibrium geometries, rotational
constants, and vibrational frequencies. In addition, we present a
method for performing dynamic damping, based on the work of Zerner
and Hehenberger on damping of the SCF equations,\cite{zernerDynamicalDampingScheme1979}
in order to combat certain difficulties experienced in Paper I.

\section{Theory}

Before reviewing the theory behind sub-iteration of the coupled cluster
(CC) amplitude equations, let us briefly recall the coupled cluster
problem using second quantization,\cite{cizek,shavitt_bartlett,helgaker}
\begin{align}
0 & =\langle\Phi_{i_{1}\ldots i_{k}}^{a_{1}\ldots a_{k}}|e^{-\hat{T}}\hat{H}_{N}e^{\hat{T}}|\Phi_{0}\rangle,\;1\le k\le M\label{eq:ccamp}\\
\hat{H}_{N}=\hat{F}_{N}+\hat{V}_{N} & =\sum\limits _{pq}f_{q}^{p}\left\{ a_{p}^{\dagger}a_{q}\right\} _{N}+\frac{1}{4}\sum\limits _{pqrs}v_{rs}^{pq}\left\{ a_{p}^{\dagger}a_{q}^{\dagger}a_{s}a_{r}\right\} _{N}\\
\hat{T}=\sum\limits _{k=1}^{M}\hat{T}_{k} & =\sum\limits _{k=1}^{M}\frac{1}{(k!)^{2}}\sum\limits _{a_{1}\ldots a_{k}}\sum\limits _{i_{1}\ldots i_{k}}t_{i_{1}\ldots i_{k}}^{a_{1}\ldots a_{k}}a_{a_{1}}^{\dagger}\ldots a_{a_{k}}^{\dagger}a_{i_{1}}\ldots a_{i_{k}}
\end{align}
where $\{\ldots\}_{N}$ denotes normal ordering. The notation and
definitions of the operators are the same as in Paper I. As previously
noted, these equations may be solved using a Jacobi-like iterative
procedure, 
\begin{align}
(f_{i_{1}}^{i_{1}}+\ldots+f_{i_{k}}^{i_{k}}-f_{a_{1}}^{a_{1}}-\ldots-f_{a_{k}}^{a_{k}})t_{i_{1}\ldots i_{k}}^{a_{1}\ldots a_{k}} & =\nonumber \\
\langle\Phi_{i_{1}\ldots i_{k}}^{a_{1}\ldots a_{k}}|e^{-\hat{T}}\left(\sum\limits _{pq}(1-\delta_{pq})f_{q}^{p}\right. & \left.\left\{ a_{p}^{\dagger}a_{q}\right\} _{N}+\hat{V}_{N}\right)e^{\hat{T}}|\Phi_{0}\rangle\\
D_{i_{1}\ldots i_{k}}^{a_{1}\ldots a_{k}}t_{i_{1}\ldots i_{k}}^{a_{1}\ldots a_{k}} & =\langle\Phi_{i_{1}\ldots i_{k}}^{a_{1}\ldots a_{k}}|e^{-\hat{T}}\hat{H}_{N}^{\prime}e^{\hat{T}}|\Phi_{0}\rangle\label{eq:jacobi}\\
\mathbf{D}_{k}\mathbf{T}_{k} & =\mathbf{Z}_{k}(\mathbf{T})
\end{align}
where the last form has been written using a tensor notation, $\mathbf{D}_{k}\equiv D_{i_{1}\ldots i_{k}}^{a_{1}\ldots a_{k}}$,
$\mathbf{T}_{k}\equiv t_{i_{1}\ldots i_{k}}^{a_{1}\ldots a_{k}}$,
etc. with $\mathbf{T}=\mathbf{T}_{1}+\mathbf{T}_{2}+\cdots$. This
basic step serves as the building block for all standard coupled cluster
methods, even when using advanced convergence acceleration techniques
such as DIIS, RLE, or CROP.

\subsection{Sub-iteration in the Amplitude Equations}

In Paper I, we introduced a modification to the Jacobi step (\ref{eq:jacobi})
where certain terms are shifted from one iteration to another such
that the converged amplitudes are equal to the usual ones (and solve
(\ref{eq:ccamp}) as required), but differ iteration-by-iteration.
In particular, we chose to ``prioritize'' contributions from the
triples and quadruples to lower-order amplitudes by using the freshly-computed
amplitudes $\mathbf{D}_{k}^{-1}\mathbf{Z}_{k}$, $k=3,4$ at a given
iteration, rather than the starting amplitudes,
\begin{eqnarray}
\mathbf{D}_{1}\mathbf{T}_{1} & = & \mathbf{Z}_{1}(\mathbf{T}_{1}+\mathbf{T}_{2}+\tilde{\mathbf{T}}_{3})\nonumber \\
 & = & \mathbf{Z}_{1}(\mathbf{T}_{1}+\mathbf{T}_{2})+\mathbf{D}_{1}\mathbf{Q}_{3,1}\label{eq:modt1}\\
\mathbf{D}_{2}\mathbf{T}_{2} & = & \mathbf{Z}_{2}(\mathbf{T}_{1}+\mathbf{T}_{2}+\tilde{\mathbf{T}}_{3}+\tilde{\mathbf{T}}_{4})\nonumber \\
 & = & \mathbf{Z}_{2}(\mathbf{T}_{1}+\mathbf{T}_{2})+\mathbf{D}_{2}\mathbf{Q}_{3,2}(\mathbf{T}_{1})+\mathbf{D}_{2}\mathbf{Q}_{4,2}\label{eq:modt2}\\
\mathbf{D}_{3}\mathbf{T}_{3} & = & \mathbf{Z}_{3}(\mathbf{T}_{1}+\mathbf{T}_{2}+\mathbf{T}_{3}+\tilde{\mathbf{T}}_{4})\nonumber \\
 & = & \mathbf{Z}_{3}(\mathbf{T}_{1}+\mathbf{T}_{2}+\mathbf{T}_{3})+\mathbf{D}_{3}\mathbf{Q}_{4,3}(\mathbf{T}_{1})\label{eq:modt3}\\
\mathbf{D}_{4}\mathbf{T}_{4} & = & \mathbf{Z}_{4}(\mathbf{T})
\end{eqnarray}
for CCSDTQ, $\tilde{\mathbf{T}}_{4}=\mathbf{D}_{4}^{-1}\mathbf{Z}_{4}(\mathbf{T})$
and $\tilde{\mathbf{T}}_{3}=\mathbf{D}_{3}^{-1}\mathbf{Z}_{3}(\mathbf{T}_{1}+\mathbf{T}_{2}+\mathbf{T}_{3}+\tilde{\mathbf{T}}_{4})$.
The corresponding form of the modified CCSDT equations simply deletes
the quadruples amplitudes and equations, while equivalent modified
equations for approximate triples and quadruples methods may be obtained
in a similar manner. The $\mathbf{Q}_{m,n}$ terms represent the prioritized
higher-order contributions, and may depend parametrically on $\hat{T}_{1}$.
Note especially that this definition of $\mathbf{Q}_{m,n}$ differs
slightly from that in Paper I by the presence of the denominators.

\begin{figure}
\begin{centering}
\includegraphics{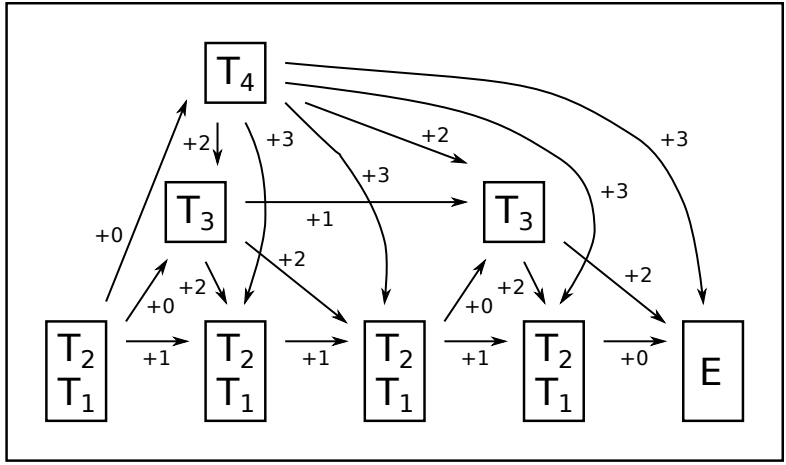}
\par\end{centering}
\caption{\label{fig:ccsdtq-sub}Illustration of the sub-iteration scheme for
the CCSDTQ amplitude equations, adapted from Ref.~\citenum{matthewsAcceleratingConvergenceHigherorder2015}.}
 
\end{figure}

Starting from this form of the amplitude equations, it is easy to
see that the lower-order equations take on the form of a standard
coupled cluster iteration, CCSD for $\hat{T}_{1}$ and $\hat{T}_{2}$
and CCSDT for $\hat{T}_{3}$, plus additional $\mathbf{Q}_{m,n}$
contributions. By holding these contributions temporarily constant,
a number of CCSD and/or CCSDT sub-iterations may be performed without
updating the highest-order amplitudes (which is also the leading-order
computational cost). In Paper I we found that this method, except
for extreme cases like BeO, was effective in both reducing the number
of iterations (defined by the number of updates to the highest-order
amplitudes) and avoiding the requirement of extrapolating the higher-order
amplitudes, e.g. using DIIS. An illustration of this sub-iteration
scheme for CCSDTQ is given in \figref{ccsdtq-sub}, where the ``relative
order'' of each contribution is given, defined as the total order
of all inputs (in the usual Møller-Plesset partitioning) minus the
(leading) order of the output. In this example, using nested sub-iteration
of the CCSD and CCSDT equations increases the relative order through
which the energy in the next iteration is correct from +1 to +3.

\subsection{Sub-iteration in the $\Lambda$ Equations}

In this work, we apply a similar analysis to the coupled cluster $\Lambda$
equations which describe the relaxation of the amplitudes with respect
to a perturbation. The solution of these equations is a necessary
prerequisite for the computation of analytic gradients and molecular
properties. As with the amplitude equations, the $\Lambda$ equations
may be written using a Jacobi-like iterative scheme,
\begin{align}
\hat{\Lambda}=\sum\limits _{k=1}^{M}\hat{\Lambda}_{k} & =\sum\limits _{k=1}^{M}\frac{1}{(k!)^{2}}\sum\limits _{a_{1}\ldots a_{k}}\sum\limits _{i_{1}\ldots i_{k}}\lambda_{a_{1}\ldots a_{k}}^{i_{1}\ldots i_{k}}a_{i_{1}}^{\dagger}\ldots a_{i_{k}}^{\dagger}a_{a_{1}}\ldots a_{a_{k}}\\
0 & =\langle\Phi_{0}|\left(1+\hat{\Lambda}\right)e^{-\hat{T}}H_{N}e^{\hat{T}}|\Phi_{i_{1}\ldots i_{k}}^{a_{1}\ldots a_{k}}\rangle,\;1\le k\le M\\
D_{i_{1}\ldots i_{k}}^{a_{1}\ldots a_{k}}\lambda_{a_{1}\ldots a_{k}}^{i_{1}\ldots i_{k}} & =\langle\Phi_{0}|\left(1+\hat{\Lambda}\right)e^{-\hat{T}}\hat{H}_{N}^{\prime}e^{\hat{T}}|\Phi_{i_{1}\ldots i_{k}}^{a_{1}\ldots a_{k}}\rangle\\
\mathbf{D}_{k}\mathbf{\Lambda}_{k} & =\mathbf{W}_{k}(\hat{T},\hat{\Lambda})
\end{align}
As in the case of the cluster amplitudes, it is straightforward to
prioritize higher-order contributions,

\begin{eqnarray}
\mathbf{D}_{1}\mathbf{\Lambda}_{1} & = & \mathbf{W}_{1}(\mathbf{T},\mathbf{\Lambda}_{1}+\mathbf{\Lambda}_{2}+\tilde{\mathbf{\Lambda}}_{3}+\tilde{\mathbf{\Lambda}}_{4})\nonumber \\
 & = & \mathbf{W}_{1}(\mathbf{T}_{1}+\mathbf{T}_{2},\mathbf{\Lambda}_{1}+\mathbf{\Lambda}_{2})+\mathbf{D}_{1}\mathbf{U}_{3,1}+\mathbf{D}_{1}\mathbf{U}_{4,1}\label{eq:modt1-1}\\
\mathbf{D}_{2}\mathbf{\Lambda}_{2} & = & \mathbf{W}_{2}(\mathbf{T},\mathbf{\Lambda}_{1}+\mathbf{\Lambda}_{2}+\tilde{\mathbf{\Lambda}}_{3}+\tilde{\mathbf{\Lambda}}_{4})\nonumber \\
 & = & \mathbf{W}_{2}(\mathbf{T}_{1}+\mathbf{T}_{2},\mathbf{\Lambda}_{1}+\mathbf{\Lambda}_{2})+\mathbf{D}_{2}\mathbf{U}_{3,2}+\mathbf{D}_{2}\mathbf{U}_{4,2}\label{eq:modt2-1}\\
\mathbf{D}_{3}\mathbf{\Lambda}_{3} & = & \mathbf{W}_{3}(\mathbf{T},\mathbf{\Lambda}_{1}+\mathbf{\Lambda}_{2}+\mathbf{\Lambda}_{3}+\tilde{\mathbf{\Lambda}}_{4})\nonumber \\
 & = & \mathbf{W}_{3}(\mathbf{T}_{1}+\mathbf{T}_{2}+\mathbf{T}_{3},\mathbf{\Lambda}_{1}+\mathbf{\Lambda}_{2}+\mathbf{\Lambda}_{3})+\mathbf{D}_{3}\mathbf{U}_{4,3}\label{eq:modt3-1}\\
\mathbf{D}_{4}\mathbf{\Lambda}_{4} & = & \mathbf{W}_{4}(\mathbf{T},\mathbf{\Lambda})
\end{eqnarray}
where $\tilde{\mathbf{\Lambda}}_{4}=\mathbf{D}_{4}^{-1}\mathbf{W}_{4}(\mathbf{T},\mathbf{\Lambda})$
and $\tilde{\mathbf{\Lambda}}_{3}=\mathbf{D}_{3}^{-1}\mathbf{W}_{3}(\mathbf{T},\mathbf{\Lambda}_{1}+\mathbf{\Lambda}_{2}+\mathbf{\Lambda}_{3}+\tilde{\mathbf{\Lambda}}_{4})$.
Note that the contributions $\mathbf{U}_{m,n}$ depend on $\hat{T}$,
but that of course these amplitudes don't change during the solution
of the $\Lambda$ equations. In particular, the $\mathbf{U}_{m,n}$
not only include the contributions from higher $\hat{\Lambda}$ amplitudes
but also from higher $\hat{T}$ amplitudes, regardless of which $\hat{\Lambda}$
they are contracted with. For example, $\mathbf{U}_{3,1}$ includes
contributions from $\mathbf{\Lambda}_{2}\mathbf{V}_{N}\mathbf{T}_{3}$.

\begin{figure}
\begin{centering}
\includegraphics{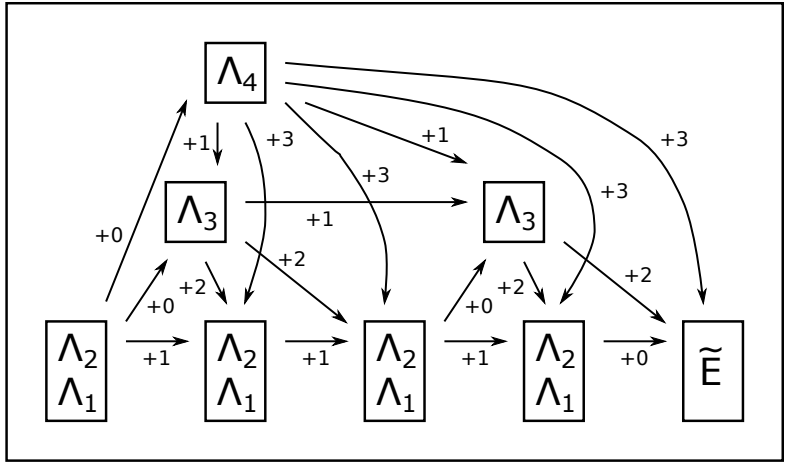}
\par\end{centering}
\caption{\label{fig:ccsdtq-lam-sub}The nested sub-iteration scheme applied
to the CCSDTQ $\Lambda$ equations.}
 
\end{figure}

Since the equations are linear in $\hat{\Lambda}$, then the $\mathbf{U}_{m,n}$
may be held constant during several modified CCSD or CCSDT $\Lambda$
iterations without approximation. Conceptually, this process is identical
to the one followed for the amplitude equations, although due to the
structure of the $\Lambda$ equations the $\mathbf{U}_{m,n}$ contributions
are rather more complicated in form. The analysis of the various contributions
from one set of amplitudes to another in terms of perturbation theory
is not quite identical, however. Most important is the fact that the
$\hat{\Lambda}_{4}$ amplitudes are second order due to the presence
of a disconnected $\hat{\Lambda}_{2}\hat{V}_{N}$ contribution, while
the other amplitudes have the same order as the corresponding cluster
amplitudes. This leads to a lowering of the relative order of the
$\hat{\Lambda}_{4}$ contributions to $\hat{\Lambda}_{3}$ by one;
but, due to the fact that the coupled cluster equations have been
solved, the ``direct'' $\hat{\Lambda}_{2}\gets\hat{\Lambda}_{4}\hat{V}_{N}$
term is cancelled out and the relative order for this contribution
remains +3. In order to analyze the convergence of the $\Lambda$
equations, let us define a ``pseudoenergy'' in analogy with the
usual coupled cluster energy,
\begin{equation}
\tilde{E}=\left(\mathbf{\Lambda}_{2}+\tfrac{1}{2}\mathbf{\Lambda}_{1}^{2}\right)\mathbf{V}_{N}+\mathbf{\Lambda}_{1}\mathbf{F}_{N}
\end{equation}
An illustration of all of the amplitude contributions for CCSDTQ to
the pseudoenergy after one ``iteration'' comprised of two CCSDT
sub-iterations and two nested CCSD micro-iterations each is given
in \figref{ccsdtq-lam-sub}. Despite the promotion of the quadruples
amplitudes from third order (in the coupled cluster equations) to
second order (in the $\Lambda$ equations), the pseudoenergy is complete
to a relative order of +3, just as in the coupled cluster equations.

While the perturbation theory analysis shows similar overall properties
for sub-iteration as applied to the $\Lambda$ equations compared
to the coupled cluster equations, the altered role of $\hat{\Lambda}_{4}$
may have some effect on the convergence properties. First, the $\hat{\Lambda}_{4}$
amplitudes may be expected to be numerically larger than the $\hat{T}_{4}$
amplitudes. Second, the coupling to the $\hat{\Lambda}_{3}$ amplitudes
is stronger. This is important as in the case of the coupled cluster
equation the effect of $\hat{T}_{4}$ on $\hat{T}_{2}$ through $\hat{T}_{3}$
is one relative order higher than the ``direct'' $\hat{T}_{2}\gets\hat{T}_{4}$
contribution. This is no longer the case in the $\Lambda$ equations
and both pathways should be expected to have a roughly equal effect.
These differences may be expected to lead to relatively slower convergence
in the $\Lambda$ case---the actual effect will be examined in \secref{Results-and-Discussion}.

\subsection{Dynamic Damping}

In Paper I, the performance of the sub-iteration scheme was severely
degraded in the case of BeO, even leading to divergence for full CCSDTQ.
An analysis of the cluster amplitudes showed that BeO has a very large
$\hat{T}_{1}$ amplitude, approximately 0.13, corresponding to rotation
of the HOMO-2 to the LUMO ($\sigma\rightarrow\sigma^{*}$). As $\hat{T}_{1}$
is normally considered second order, with numerical values below 0.01
or so, this violation of the usual Møller-Plesset picture was thought
to lead to this breakdown. However, a reanalysis of the relative orders
with reassignment of $\hat{T}_{1}$ as first order, or even zeroth
order, does not change the prescribed structure of the sub-iteration
scheme. Essentially, this is due to the fact that $\hat{T}_{1}$,
because it must always be contracted with the Hamiltonian, is never
solely responsible for a change in the excitation rank. Rather, it
``dresses'' the Hamiltonian as already present. Thus, the leading
order terms still contain the same structure with respect to the Hamiltonian
and the higher-order amplitudes.

Instead, we can turn to an analysis of the convergence (or rather
non-convergence) pattern of the energy and amplitudes. In the BeO
case, it is seen that the amplitude equations, starting as early as
the fifth iteration, enter into a strongly oscillatory pattern when
sub-iteration is used. Our hypothesis is that the numerically-large
contributions from $\hat{T}_{1}$ cause a systematic ``overshoot''
of the amplitude updates in (\ref{eq:jacobi}) and its modified form.
This oscillatory behavior immediately recommends the use of damping,
and we found that damping of the highest-order amplitudes ($\hat{T}_{3}$
for CCSDT and $\hat{T}_{4}$ for CCSDTQ) by a factor of 0.5 leads
to a recovery of convergence for BeO. However, the use of a static
damping parameter is highly inconvenient as it precludes global use
(since damping in non-oscillatory cases will simply slow down convergence
and negate the benefit of sub-iteration). Additionally, the simple
choice of 0.5 is not necessarily the correct value to recover convergence
in other possible cases with large $\hat{T}_{1}$ amplitudes or other
sources of oscillatory behavior.

To facilitate a robust solution, we turn to the literature on damping
in the SCF method, in particular the work of Zerner and Hehenberger.\cite{zernerDynamicalDampingScheme1979}
They define derived quantities $A_{n}(in)$, $A_{n}(out)$, and $A_{n}^{*}$
with the relationships,
\begin{align}
A_{n+1}(out) & =F\left(A_{n}(in)\right)\\
A_{n+1}^{*}=A_{n+1}(in) & =(1-\alpha)A_{n+1}(out)+\alpha A_{n}(in)\label{eq:alpha}\\
A_{n+1}(out) & =A_{n}(in)\;\text{as}\;n\rightarrow\infty\label{eq:damp}
\end{align}
where $F$ is a mapping function specific to the iterative method;
in the SCF case this is the solution of the Roothaan-Hall equations.
The quantity $A$ depends on the parameters of the iterative procedure,
e.g. the density matrix to continue the SCF example. Zerner and Hehenberger
suggested the use of the average Mulliken populations, since these
are sensitive to changes in charge distribution and hence give a detailed
measurement of the rate of change of the density. In order to find
an ideal damping parameter $\alpha$, the self-consistent solution
(\ref{eq:damp}) can be used to define an extrapolation of the line,
\begin{align}
A_{\infty}(out)-A_{n}(out) & =m\left(A_{\infty}(in)-A_{n-1}(in)\right)\\
m & \approx\frac{A_{n+1}(out)-A_{n}(out)}{A_{n}(in)-A_{n-1}(in)}
\end{align}
to a point on the diagonal ($A_{\infty}(out)=A_{\infty}(in)$),
\begin{align}
A_{n+1}^{*}=A_{n+1}(in) & =\frac{A_{n+1}(out)-mA_{n}(in)}{1-m}\\
\alpha & =\frac{m}{m-1}
\end{align}
where $\alpha$ is set to zero if $m\ge0$, as this would correspond
to extrapolation rather than damping.

To apply this idea to the coupled cluster and $\Lambda$ equations,
the $F$ function now becomes the Jacobi-like iterative step which
produces a new set of amplitudes from a starting set. At convergence,
of course, this step is self-consistent. Now the remaining difficulty
is the choice of an effective $A$ measure. Initial experimentation
with the norm of the change in the iterations, $A(\mathbf{T})=\Vert\mathbf{T}-\mathbf{T}_{old}\Vert_{2}$,
was not effective. In retrospect, this is not unexpected, as during
oscillatory behavior the \emph{norm} of the change in the amplitudes
is relatively constant, while the \emph{sign} flips each iteration.
Thus, an effective $A$ value must include the sign as well as the
magnitude of amplitude changes. Instead, we arrived at the definition
of $A$ in terms of partial energies,
\begin{align}
A(\mathbf{T}_{3}) & =\mathbf{V}_{N}\left(\mathbf{Q}_{3,2}+\tfrac{1}{2}\mathbf{Q}_{3,1}^{2}\right)+\mathbf{F}_{N}\mathbf{Q}_{3,1}\\
A(\mathbf{T}_{4}) & =\mathbf{V}_{N}\left(\mathbf{Q}_{4,2}+\mathbf{Q}_{4,3,2}+\tfrac{1}{2}\mathbf{Q}_{4,3,1}^{2}\right)+\mathbf{F}_{N}\mathbf{Q}_{4,3,1}
\end{align}
where the new quantities $\mathbf{Q}_{m,n,k}$ represent the chained
contribution of $\hat{T}_{m}$ to $\hat{T}_{k}$ through $\hat{T}_{n}$,
\begin{align}
\mathbf{Q}_{4,3,1} & =\mathbf{D}_{1}^{-1}\mathbf{V}_{N}\mathbf{Q}_{4,3}\\
\mathbf{Q}_{4,3,2} & =\mathbf{D}_{2}^{-1}\left(\mathbf{V}_{N}+\mathbf{V}_{N}\mathbf{T}_{1}\right)\mathbf{Q}_{4,3}
\end{align}
This approach proved much more fruitful as will be seen in \secref{Results-and-Discussion}.

Application of dynamic damping to the $\Lambda$ equations can be
achieved by using a similar definition of $A$ based on the pseudoenergy,
\begin{align}
A(\mathbf{\Lambda}_{3}) & =\left(\mathbf{U}_{3,2}+\tfrac{1}{2}\mathbf{U}_{3,1}^{2}\right)\mathbf{V}_{N}+\mathbf{U}_{3,1}\mathbf{F}_{N}\\
A(\mathbf{\Lambda}_{4}) & =\left(\mathbf{U}_{4,2}+\mathbf{U}_{4,3,2}+\tfrac{1}{2}\mathbf{U}_{4,1}^{2}+\tfrac{1}{2}\mathbf{U}_{4,3,1}^{2}\right)\mathbf{V}_{N}+\left(\mathbf{U}_{4,1}+\mathbf{U}_{4,3,1}\right)\mathbf{F}_{N}
\end{align}
The $\mathbf{U}_{4,3,1}$ and $\mathbf{U}_{4,3,2}$ contributions
are considerably more complicated than $\mathbf{Q}_{4,3,1}$ and $\mathbf{Q}_{4,3,2}$
and we do not reproduce them in full here, although they may be easily
calculated using parts of the standard $\Lambda$ equations. Due to
computational efficiency concerns detailed in the next section, we
do not actually use the full $\mathbf{U}_{m,n}$ contributions, rather
we include only those parts that can be computed at a cost of $\mathscr{O}(N^{7})$
or lower for CCSDT or $\mathscr{O}(N^{9})$ or lower for CCSDTQ, where
$N$ is a measure of the system size.

\subsection{\label{subsec:Implementation-Details}Implementation Details}

The application of sub-iteration to the coupled cluster equations
is, in theory, rather straightforward, as is the use of dynamic damping.
For the $\Lambda$ equations, the equations become significantly more
complicated, but the essential technique is unchanged. In reality,
though, there are a number of subtle issues that can drastically affect
the performance of the method (measured by the rate of convergence),
as well as the computational efficiency. We provide a list of issues
that we encountered during the implementation of the methods presented,
along with our recommended resolution or implementation guidance,
in the hopes that others may avoid similar pitfalls and to increase
the reproducibility of our results.
\begin{enumerate}
\item \figref{ccsdtq-sub} suggests that the structure of a complete iteration
(including sub- and micro-iterations) is that a complete CCSDTQ iteration
should be performed first, followed by a modified CCSDT iteration,
followed by CCSD, etc. However, in practice we have found that performing
these steps in reverse order is the most stable. That is, for example
in an iteration with 2 CCSD micro-iterations for each of 2 CCSDT sub-iterations,
one should perform a sequence of steps (a)-(a)-(b)-(a)-(a)-(b)-(a)-(a)-(c),
where (a) is a CCSD micro-iteration, (b) is a CCSDT sub-iteration,
and (c) is a full CCSDTQ iteration. When using DIIS acceleration of
the $\hat{T}_{1}$ and $\hat{T}_{2}$ (and possibly $\hat{T}_{3}$)
amplitudes, putting the full iteration last provides a more self-consistent
picture of the current amplitudes and leads to more a effective use
of the extrapolation. The same observation is true of the $\Lambda$
equations.
\item When using DIIS on the lower-order amplitudes, e.g. extrapolating
only $\hat{T}_{1}$ and $\hat{T}_{2}$ in a CCSDTQ calculation, it
is tempting to apply DIIS at every micro-iteration. We have found
that this is not effective, and often actively detrimental to timely
convergence.
\item In a CCSDTQ calculation without extrapolation of $\hat{T}_{3}$, we
have found that allowing damping of both $\hat{T}_{3}$ and $\hat{T}_{4}$
using the above procedure is often prone to runaway behavior. We hypothesize
that this is because the CCSDT sub-iterations immediately following
an update of $\hat{T}_{4}$ and those in subsequent relaxation of
the lower-order amplitudes produce qualitatively distinct changes
in the $\hat{T}_{3}$ amplitudes which are not compatible with a geometric
extrapolation scheme. In any case, we have found that dynamic damping
applied only to the $\hat{T}_{4}$ amplitudes is sufficient to recover
convergence even in very difficult cases.
\item The calculation of the $A$ measure by its definition requires construction
of the $\mathbf{Q}_{m,n}$ or $\mathbf{U}_{m,n}$ tensors. Using the
Zerner-Hehenberger scheme, $A$ must be computed twice each iteration:
once using the initial amplitudes and once after the Jacobi update.
When damping is applied, $\mathbf{Q}_{m,n}$/$\mathbf{U}_{m,n}$ for
the final damped amplitudes must also be constructed. This could be
accomplished by saving the ``before'' and ``after'' versions and
combining them as in (\ref{eq:alpha}), but we chose instead to recalculate
the contributions in this case.
\item Closely related to the last point, the calculation of $\mathbf{Q}_{m,n}$
(and hence $A(\mathbf{T})$) has a scaling of $\mathscr{O}(N^{7})$
for CCSDT and $\mathscr{O}(N^{9})$ for CCSDTQ. While this scaling
is lower than the full CCSDT or CCSDTQ iteration, it is often a not-insignificant
cost. For $\mathbf{U}_{m,n}$ (and $A(\mathbf{\Lambda})$), though,
a full computation requires $\mathscr{O}(N^{8})$ and $\mathscr{O}(N^{10})$
cost for CCSDT and CCSDTQ, respectively. While this high cost is necessary
for $\mathbf{U}_{m,n}$ in order to maintain correctness, we choose
to omit the highest-scaling terms in the computation of $A(\mathbf{\Lambda})$.
This is especially important as $A(\mathbf{\Lambda})$ has to be computed
twice, along with an additional (re-)computation of $\mathbf{U}_{m,n}$
if damping is applied. This choice maintains the number of highest-scaling
steps to be the same as in the original method.
\end{enumerate}

\section{\label{sec:Results-and-Discussion}Results and Discussion}

The sub-iteration method for the coupled cluster $\Lambda$ equations
and the proposed dynamic damping scheme have been implemented in a
development version of the CFOUR program suite.\cite{CFOUR} As in
Paper I, we evaluate the performance of these methods by performing
frozen-core calculations on $\ce{H2O}$, $\ce{BeO}$, $\ce{C2}$,
and $\ce{O3}$. For the first three molecules, we use the cc-pVQZ
basis set for triples methods (CCSDT and its iterative approximations
CCSDT-1a,\cite{ccsdt-1a} CCSDT-1b,\cite{nogaFullCCSDTModel1987}
CCSDT-2,\cite{nogaFullCCSDTModel1987} CCSDT-3,\cite{nogaFullCCSDTModel1987}
and CC3\cite{cc3}) and cc-pVTZ for quadruples methods (CCSDTQ and
its approximations\cite{kallayApproximateTreatmentHigher2005} CCSDTQ-1a,
CCSDTQ-1b, CCSDT-3, and CC4). For ozone, smaller basis sets cc-pVTZ
and cc-pVDZ were used, respectively. Unless otherwise noted, the calculations
are considered converged when $\max(\Vert\Delta\mathbf{T}_{1}\Vert_{\infty},\Vert\Delta\mathbf{T}_{2}\Vert_{\infty})<10^{-7}$.
Lastly, when referring to a specific instantiation of the sub-iteration
method, we refer to CCSDT$_{m}$, CC3$_{m}$, CCSDTQ$_{m,n}$, CCSDTQ-3$_{m,n}$,
etc. as defined in Paper I. The sub-iteration methods are compared
to calculations using full DIIS (where all $\hat{T}$ amplitudes are
extrapolated together). In all cases, five DIIS expansion vectors
are used.

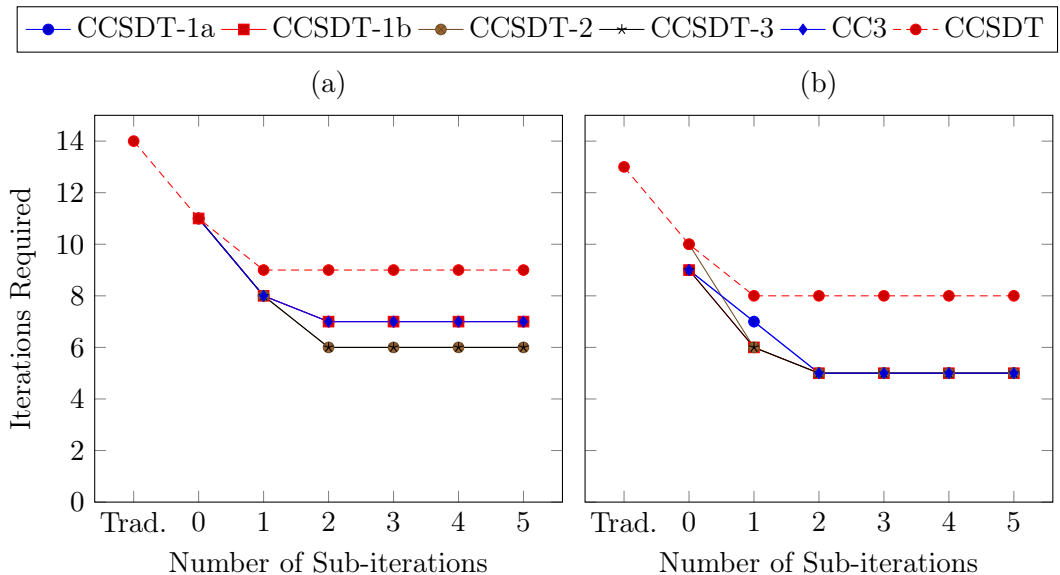
\begin{figure}
\begin{tikzpicture}
\pgfplotsset{try min ticks=6}
\begin{groupplot}[
width=7.8cm,
group style={y descriptions at=edge left,group size=2 by 1,group name=plots,horizontal sep=0.3cm},
ymin=0,
ymax=15,
ylabel near ticks,
symbolic x coords = {Trad., 0, 1, 2, 3, 4, 5},
xlabel=Number of Sub-iterations,
ylabel={Iterations Required}
]

\nextgroupplot[
legend to name=legend,
legend columns=-1]
\legend{CCSDT-1a, CCSDT-1b, CCSDT-2, CCSDT-3, CC3, CCSDT}
\addplot +[ sharp plot ] coordinates {
(0, 11)
(1, 8)
(2, 7)
(3, 7)
(4, 7)
(5, 7)};
\addplot +[ sharp plot ] coordinates {
(0, 11)
(1, 8)
(2, 7)
(3, 7)
(4, 7)
(5, 7)};
\addplot +[ sharp plot ] coordinates {
(0, 11)
(1, 8)
(2, 6)
(3, 6)
(4, 6)
(5, 6)};
\addplot +[ sharp plot ] coordinates {
(0, 11)
(1, 8)
(2, 6)
(3, 6)
(4, 6)
(5, 6)};
\addplot +[ sharp plot ] coordinates {
(0, 11)
(1, 8)
(2, 7)
(3, 7)
(4, 7)
(5, 7)};
\addplot +[ sharp plot ] coordinates {
(Trad., 14)
(0, 11)
(1, 9)
(2, 9)
(3, 9)
(4, 9)
(5, 9)};

\nextgroupplot
\addplot +[ sharp plot ] coordinates {
(0, 9)
(1, 7)
(2, 5)
(3, 5)
(4, 5)
(5, 5)};
\addplot +[ sharp plot ] coordinates {
(0, 9)
(1, 6)
(2, 5)
(3, 5)
(4, 5)
(5, 5)};
\addplot +[ sharp plot ] coordinates {
(0, 10)
(1, 6)
(2, 5)
(3, 5)
(4, 5)
(5, 5)};
\addplot +[ sharp plot ] coordinates {
(0, 9)
(1, 6)
(2, 5)
(3, 5)
(4, 5)
(5, 5)};
\addplot +[ sharp plot ] coordinates {
(0, 9)
(1, 7)
(2, 5)
(3, 5)
(4, 5)
(5, 5)};
\addplot +[ sharp plot ] coordinates {
(Trad., 13)
(0, 10)
(1, 8)
(2, 8)
(3, 8)
(4, 8)
(5, 8)};

\end{groupplot}

\node at (plots c1r1.north east) [inner sep=0pt,anchor=south, yshift=5ex,xshift=-2ex] {\ref{legend}};
\draw([yshift=0.4cm]plots c1r1.north) circle (0pt) node {(a)};
\draw([yshift=0.4cm]plots c2r1.north) circle (0pt) node {(b)};

\end{tikzpicture}\caption{\label{fig:h2o-ccsdt}Number of full iterations required to convergence
the CC amplitude equations (a) and $\Lambda$ equations (b) for $\ce{H2O}$,
with varying numbers of CCSD sub-iterations. For CCSDT, zero sub-iterations
indicates only prioritization of higher-order contributions, while
``Trad.'' indicates a standard CC calculation. For the approximate
methods, ``Trad.'' and zero sub-iterations are identical.}
\end{figure}

First, we examine the behavior of the sub-iteration method applied
to the $\Lambda$ equations in the ``well-behaved'' case of $\ce{H2O}$
and compare to that previously observed for the amplitude equations.
It should be noted, though, that due to ongoing work on the sub-iteration
method in general, and our practical experiences as summarized in
\subsecref{Implementation-Details}, that the results for the amplitude
equations presented here are not identical to those in Paper I. The
main observations in that paper still hold, however. \figref{h2o-ccsdt}
summarizes the effect of sub-iterations on the triples methods. As
the number of sub-iterations is increased, the number of iterations
required to reach convergence decreases for both sets of equations,
and appears to plateau after two sub-iterations. In the case of the
$\Lambda$ equations, it seems that slightly fewer iterations are
needed in both the traditional and sub-iteration-accelerated cases
compared to the amplitude equations. Here, we include several approximate
triples methods not considered in Paper I, in particular CCSDT-2 and
CCSDT-3. These methods include terms non-linear in $\hat{T}_{2}$
in the triples amplitude equations. When combined with sub-iteration,
the convergence rate is slightly higher (one fewer iterations is required
in this experiment). This might be an effect of the increased importance
of $\hat{T}_{2}$ relaxation via sub-iteration due to the increased
non-linearity. In general, though, it seems that a choice of 2-3 or
more CCSD sub-iterations is sufficient to gain the full benefit of
sub-iteration. Including a much larger number of sub-iterations would
lead to increased computational cost.

\begin{figure}
\begin{tikzpicture}
\pgfplotsset{try min ticks=6}
\begin{groupplot}[
width=7.8cm,
ylabel near ticks,
group style={y descriptions at=edge left,group size=2 by 2,group name=plots2,horizontal sep=0.3cm,vertical sep=1.8cm},
ymin=0,
ymax=18,
symbolic x coords = {Trad., 0, 1, 2, 3, 4, 5},
ylabel={Iterations Required}
]

\nextgroupplot[
xlabel=Number of Sub-iterations,
legend to name=legend,
legend columns=-1]
\legend{CCSDTQ-1a, CCSDTQ-1b, CCSDTQ-3, CC4, CCSDTQ}
\addplot +[ sharp plot ] coordinates {
(Trad., 14)
(0, 12)
(1, 6)
(2, 5)
(3, 5)
(4, 4)
(5, 4)};
\addplot +[ sharp plot ] coordinates {
(Trad., 14)
(0, 11)
(1, 6)
(2, 6)
(3, 5)
(4, 5)
(5, 5)};
\addplot +[ sharp plot ] coordinates {
(Trad., 14)
(0, 11)
(1, 6)
(2, 5)
(3, 5)
(4, 5)
(5, 5)};
\addplot +[ sharp plot ] coordinates {
(Trad., 14)
(0, 11)
(1, 6)
(2, 6)
(3, 5)
(4, 5)
(5, 5)};
\addplot +[ sharp plot ] coordinates {
(Trad., 17)
(0, 11)
(1, 6)
(2, 7)
(3, 6)
(4, 6)
(5, 6)};

\nextgroupplot[xlabel=Number of Sub-iterations]
\addplot +[ sharp plot ] coordinates {
(Trad., 13)
(0, 10)
(1, 6)
(2, 5)
(3, 4)
(4, 4)
(5, 3)};
\addplot +[ sharp plot ] coordinates {
(Trad., 13)
(0, 10)
(1, 6)
(2, 5)
(3, 5)
(4, 5)
(5, 5)};
\addplot +[ sharp plot ] coordinates {
(Trad., 13)
(0, 10)
(1, 6)
(2, 5)
(3, 5)
(4, 5)
(5, 5)};
\addplot +[ sharp plot ] coordinates {
(Trad., 13)
(0, 10)
(1, 6)
(2, 5)
(3, 5)
(4, 5)
(5, 5)};
\addplot +[ sharp plot ] coordinates {
(Trad., 17)
(0, 10)
(1, 8)
(2, 7)
(3, 6)
(4, 6)
(5, 6)};

\nextgroupplot[xlabel={Number of Sub- and Micro-iterations}]
\addplot +[ sharp plot ] coordinates {
(Trad., 14)
(0, 11)
(1, 6)
(2, 5)
(3, 5)
(4, 5)
(5, 5)};
\addplot +[ sharp plot ] coordinates {
(Trad., 14)
(0, 11)
(1, 6)
(2, 5)
(3, 5)
(4, 5)
(5, 5)};
\addplot +[ sharp plot ] coordinates {
(Trad., 14)
(0, 10)
(1, 6)
(2, 5)
(3, 4)
(4, 4)
(5, 4)};
\addplot +[ sharp plot ] coordinates {
(Trad., 14)
(0, 11)
(1, 6)
(2, 5)
(3, 5)
(4, 5)
(5, 5)};
\addplot +[ sharp plot ] coordinates {
(Trad., 17)
(0, 11)
(1, 6)
(2, 6)
(3, 6)
(4, 6)
(5, 6)};

\nextgroupplot[xlabel={Number of Sub- and Micro-iterations}]
\addplot +[ sharp plot ] coordinates {
(Trad., 13)
(0, 9)
(1, 5)
(2, 3)
(3, 3)
(4, 3)
(5, 3)};
\addplot +[ sharp plot ] coordinates {
(Trad., 13)
(0, 9)
(1, 5)
(2, 5)
(3, 5)
(4, 5)
(5, 5)};
\addplot +[ sharp plot ] coordinates {
(Trad., 13)
(0, 9)
(1, 5)
(2, 4)
(3, 5)
(4, 5)
(5, 5)};
\addplot +[ sharp plot ] coordinates {
(Trad., 13)
(0, 9)
(1, 5)
(2, 5)
(3, 5)
(4, 5)
(5, 5)};
\addplot +[ sharp plot ] coordinates {
(Trad., 17)
(0, 9)
(1, 6)
(2, 6)
(3, 6)
(4, 6)
(5, 6)};

\end{groupplot}

\node at (plots2 c1r1.north east) [inner sep=0pt,anchor=south, yshift=5ex,xshift=-2ex] {\ref{legend}};
\draw([yshift=0.4cm]plots2 c1r1.north) circle (0pt) node {(a)};
\draw([yshift=0.4cm]plots2 c2r1.north) circle (0pt) node {(b)};
\draw([yshift=0.4cm]plots2 c1r2.north) circle (0pt) node {(c)};
\draw([yshift=0.4cm]plots2 c2r2.north) circle (0pt) node {(d)};

\end{tikzpicture}\caption{\label{fig:h2o-ccsdtq}Number of full iterations required to convergence
the CC amplitude equations (a,c) and $\Lambda$ equations (b,d) for
$\ce{H2O}$. In (a) and (b), $\hat{T}_{1}$, $\hat{T}_{2}$, and $\hat{T}_{3}$
(or the equivalent $\hat{\Lambda}$ amplitudes) are included in the
DIIS extrapolation and the number of CCSDT sub-iterations is given.
In (c) and (d), only $\hat{T}_{1}$ and $\hat{T}_{2}$ (or $\hat{\Lambda}_{1}$
and $\hat{\Lambda}_{2}$) are extrapolated, and an equal number of
CCSDT sub-iterations and CCSD micro-iterations are used as indicated.
In all cases, zero sub-iterations indicates only prioritization of
higher-order contributions, while ``Trad.'' indicates a standard
CC calculation.}
\end{figure}
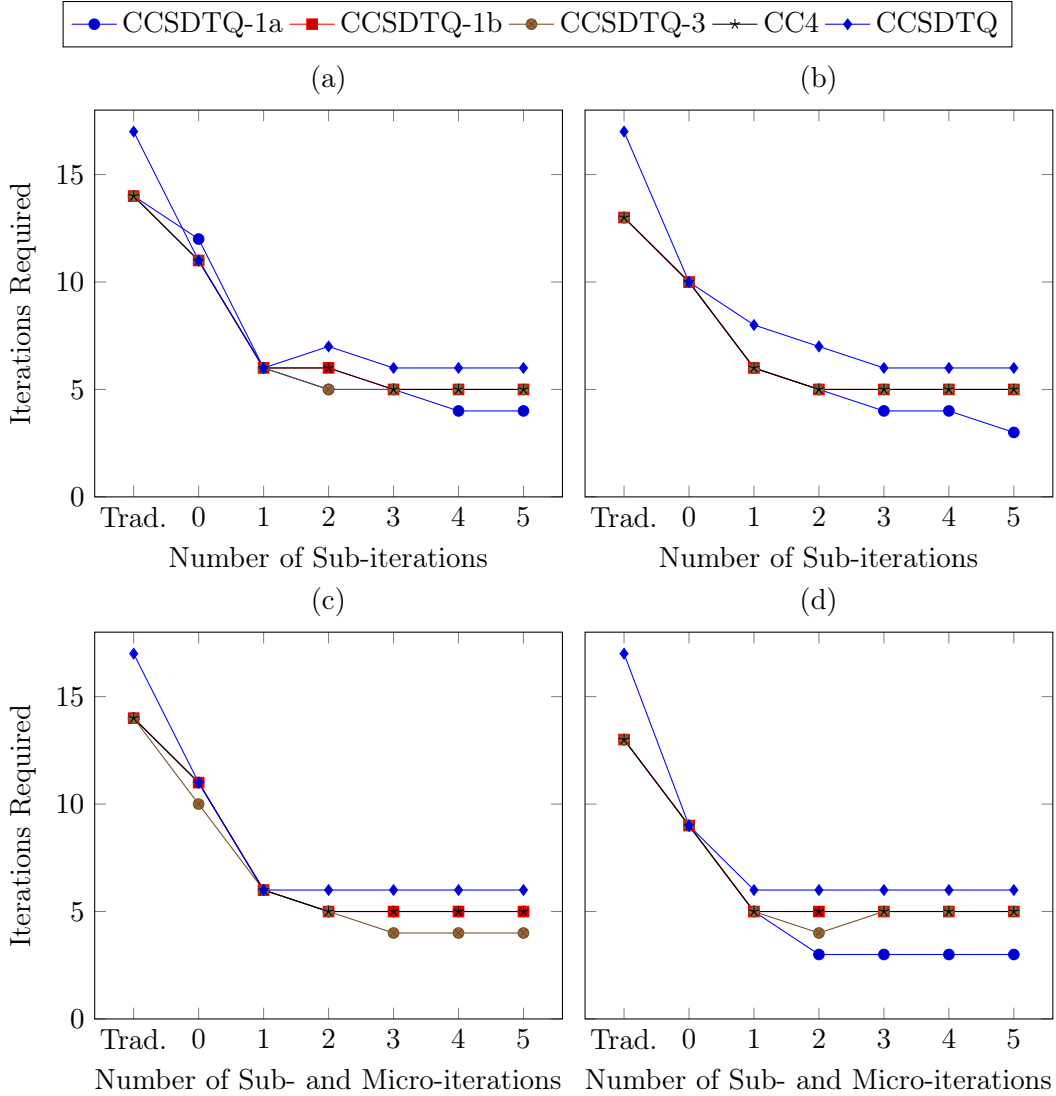

Next, a similar comparison is made for the quadruples methods, as
summarized in \figref{h2o-ccsdtq}. In this case, there is additional
flexibility in the space of possible sub-iteration methods, since
CCSDT sub- and CCSD micro-iterations may be combined. Also, one may
forgo CCSD micro-iteration in favor of extrapolating the $\hat{T}_{3}$
(or $\hat{\Lambda}_{3}$). In Paper I, it was determined that methods
of the type CCSDTQ$_{m,0}$ or CCSDTQ$_{0,n}$ (without extrapolation
of $\hat{T}_{3}$) were not as effective as combining sub-iteration
at both the triples and singles/doubles level. Thus, we only compare
CCSDTQ$_{m,m}$ methods with ``CCSDTQ$_{-,m}$'', i.e. the triples-extrapolated
approach described above). For both the amplitude and $\Lambda$ equations,
the drop in iterations required for convergence with increasing sub-iteration
is more rapid and consistent for CCSDTQ$_{m,m}$. This is especially
true for CCSDTQ, which is perhaps the method in most need of convergence
acceleration due to its extremely high computational cost. Because
CCSDTQ$_{m,m}$ not only seems more effective in accelerating convergence,
but is more economical in terms of memory and/or disk I/O compared
to CCSDTQ$_{-,m}$, we recommend the former. In contrast to the triples
methods, the amplitude and $\Lambda$ equations show extremely similar
convergence behavior, with the exception of CCSDTQ-1a. This method
does not include any direct coupling of $\hat{\Lambda}_{4}$ to $\hat{\Lambda}_{3}$,
and this simpler structure may contribute to the extremely small number
of iterations required for convergence with sub-iteration. As with
the triples methods, 2-3 sub- and micro-iterations seem sufficient
to gain the full benefit. These results support the choice of the
default settings in our CFOUR implementation of 2 CCSDT sub-iterations
and 3 CCSD sub- or micro-iterations. These settings are used for all
of the subsequent calculations.

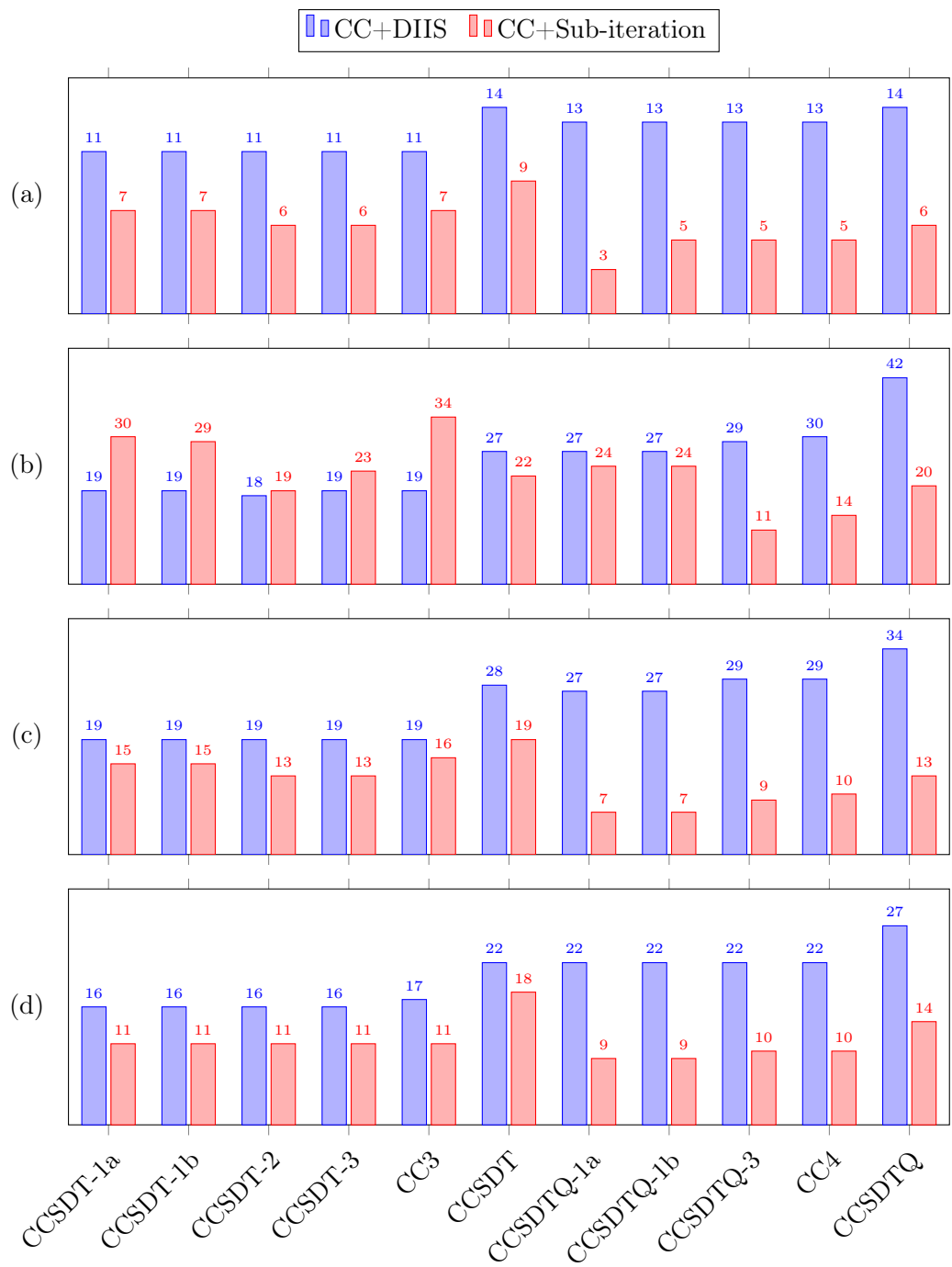
\begin{figure}
\begin{tikzpicture}
\begin{groupplot}[
group style={group size=1 by 4, vertical sep=0.5cm, group name=plots3},
ybar,
ymin=0,
width=\textwidth,
height=5cm,
xtick=data,
ytick=\empty,
enlarge x limits=.05,
nodes near coords style={font=\tiny},
nodes near coords,
nodes near coords align={vertical},
symbolic x coords = {CCSDT-1a, CCSDT-1b, CCSDT-2, CCSDT-3, CC3, CCSDT, CCSDTQ-1a, CCSDTQ-1b, CCSDTQ-3, CC4, CCSDTQ}
]

\nextgroupplot[
bar width=10,
xticklabels={},
ymax=16,
legend columns=4,
legend style={/tikz/every even column/.append style={column sep=0.25cm}},
legend style={overlay,at={(0.5,1.2)},anchor=center}
]
\addplot coordinates {
(CCSDT-1a, 11)
(CCSDT-1b, 11)
(CCSDT-2, 11)
(CCSDT-3, 11)
(CC3, 11)
(CCSDT, 14)
(CCSDTQ-1a,  13)
(CCSDTQ-1b,  13)
(CCSDTQ-3,  13)
(CC4,  13)
(CCSDTQ, 14)};
\addplot coordinates {
(CCSDT-1a, 7)
(CCSDT-1b, 7)
(CCSDT-2, 6)
(CCSDT-3, 6)
(CC3, 7)
(CCSDT, 9)
(CCSDTQ-1a, 3)
(CCSDTQ-1b,  5)
(CCSDTQ-3,  5)
(CC4,  5)
(CCSDTQ, 6)};
\legend{CC+DIIS, CC+Sub-iteration}

\nextgroupplot[
bar width=10,
xticklabels={},
ymax=48
]
\addplot coordinates {
(CCSDT-1a, 19)
(CCSDT-1b, 19)
(CCSDT-2, 18)
(CCSDT-3, 19)
(CC3, 19)
(CCSDT, 27)
(CCSDTQ-1a,  27)
(CCSDTQ-1b,  27)
(CCSDTQ-3,  29)
(CC4,  30)
(CCSDTQ, 42)};
\addplot coordinates {
(CCSDT-1a, 30)
(CCSDT-1b, 29)
(CCSDT-2, 19)
(CCSDT-3, 23)
(CC3, 34)
(CCSDT, 22)
(CCSDTQ-1a, 24)
(CCSDTQ-1b, 24)
(CCSDTQ-3, 11)
(CC4, 14)
(CCSDTQ, 20)};

\nextgroupplot[
bar width=10,
xticklabels={},
ymax=39
]
\addplot coordinates {
(CCSDT-1a, 19)
(CCSDT-1b, 19)
(CCSDT-2, 19)
(CCSDT-3, 19)
(CC3, 19)
(CCSDT, 28)
(CCSDTQ-1a,  27)
(CCSDTQ-1b,  27)
(CCSDTQ-3,  29)
(CC4,  29)
(CCSDTQ, 34)};
\addplot coordinates {
(CCSDT-1a, 15)
(CCSDT-1b, 15)
(CCSDT-2, 13)
(CCSDT-3, 13)
(CC3, 16)
(CCSDT, 19)
(CCSDTQ-1a, 7)
(CCSDTQ-1b,  7)
(CCSDTQ-3,  9)
(CC4,  10)
(CCSDTQ, 13)};

\nextgroupplot[
bar width=10,
ymax=32,
x tick label style={rotate=45,anchor=north east}
]
\addplot coordinates {
(CCSDT-1a, 16)
(CCSDT-1b, 16)
(CCSDT-2, 16)
(CCSDT-3, 16)
(CC3, 17)
(CCSDT, 22)
(CCSDTQ-1a,  22)
(CCSDTQ-1b,  22)
(CCSDTQ-3,  22)
(CC4,  22)
(CCSDTQ, 27)};
\addplot coordinates {
(CCSDT-1a, 11)
(CCSDT-1b, 11)
(CCSDT-2, 11)
(CCSDT-3, 11)
(CC3, 11)
(CCSDT, 18)
(CCSDTQ-1a, 9)
(CCSDTQ-1b,  9)
(CCSDTQ-3,  10)
(CC4,  10)
(CCSDTQ, 14)};

\end{groupplot}

\draw([xshift=-0.6cm]plots3 c1r1.west) circle (0pt) node {(a)};
\draw([xshift=-0.6cm]plots3 c1r2.west) circle (0pt) node {(b)};
\draw([xshift=-0.6cm]plots3 c1r3.west) circle (0pt) node {(c)};
\draw([xshift=-0.6cm]plots3 c1r4.west) circle (0pt) node {(d)};

\end{tikzpicture}

\caption{\label{fig:all-cc}Comparison of the number of iterations required
for convergence in the CC amplitude equations for full DIIS and the
suggested sub-iteration method (2 CCSDT sub-iterations and 3 CCSD
sub- or micro-iterations). Dynamic damping is included in the sub-iteration
CCSDT and CCSDTQ calculations. (a) $\ce{H2O}$, (b) $\ce{BeO}$, (c)
$\ce{O3}$, (d) $\ce{C2}$.}
\end{figure}

\begin{figure}
\begin{tikzpicture}
\begin{groupplot}[
group style={group size=1 by 4, vertical sep=0.5cm, group name=plots4},
ymin=0,
ybar,
width=\textwidth,
height=5cm,
xtick=data,
ytick=\empty,
enlarge x limits=.05,
nodes near coords style={font=\tiny},
nodes near coords,
nodes near coords align={vertical},
symbolic x coords = {CCSDT-1a, CCSDT-1b, CCSDT-2, CCSDT-3, CC3, CCSDT, CCSDTQ-1a, CCSDTQ-1b, CCSDTQ-3, CC4, CCSDTQ}
]

\nextgroupplot[
bar width=10,
xticklabels={},
ymax=15,
legend columns=4,
legend style={/tikz/every even column/.append style={column sep=0.25cm}},
legend style={overlay,at={(0.5,1.2)},anchor=center}
]
\addplot coordinates {
(CCSDT-1a, 9)
(CCSDT-1b, 9)
(CCSDT-2, 10)
(CCSDT-3, 9)
(CC3, 9)
(CCSDT, 12)
(CCSDTQ-1a, 12)
(CCSDTQ-1b,  11)
(CCSDTQ-3,  11)
(CC4,  11)
(CCSDTQ, 13)};
\addplot coordinates {
(CCSDT-1a, 5)
(CCSDT-1b, 5)
(CCSDT-2, 5)
(CCSDT-3, 5)
(CC3, 5)
(CCSDT, 8)
(CCSDTQ-1a, 3)
(CCSDTQ-1b,  5)
(CCSDTQ-3,  4)
(CC4,  5)
(CCSDTQ, 6)};
\legend{$\Lambda$+DIIS, $\Lambda$+Sub-iteration}

\nextgroupplot[
bar width=10,
xticklabels={},
ymax=37
]
\addplot coordinates {
(CCSDT-1a, 18)
(CCSDT-1b, 17)
(CCSDT-2, 18)
(CCSDT-3, 17)
(CC3, 17)
(CCSDT, 32)
(CCSDTQ-1a, 31)
(CCSDTQ-1b, 31)
(CCSDTQ-3, 26)
(CC4, 24)
(CCSDTQ, 29)};
\addplot coordinates {
(CCSDT-1a, 27)
(CCSDT-1b, 26)
(CCSDT-2, 18)
(CCSDT-3, 20)
(CC3, 30)
(CCSDT, 22)
(CCSDTQ-1a, 27)
(CCSDTQ-1b, 27)
(CCSDTQ-3, 13)
(CC4, 16)
(CCSDTQ, 19)};

\nextgroupplot[
bar width=10,
xticklabels={},
ymax=33
]
\addplot coordinates {
(CCSDT-1a, 20)
(CCSDT-1b, 17)
(CCSDT-2, 18)
(CCSDT-3, 17)
(CC3, 18)
(CCSDT, 24)
(CCSDTQ-1a, 24)
(CCSDTQ-1b,  24)
(CCSDTQ-3,  21)
(CC4,  24)
(CCSDTQ, 29)};
\addplot coordinates {
(CCSDT-1a, 14)
(CCSDT-1b, 14)
(CCSDT-2, 12)
(CCSDT-3, 12)
(CC3, 14)
(CCSDT, 23)
(CCSDTQ-1a, 6)
(CCSDTQ-1b,  6)
(CCSDTQ-3,  10)
(CC4,  10)
(CCSDTQ, 15)};

\nextgroupplot[
bar width=10,
ymax=32,
x tick label style={rotate=45,anchor=north east}
]
\addplot coordinates {
(CCSDT-1a, 18)
(CCSDT-1b, 19)
(CCSDT-2, 18)
(CCSDT-3, 19)
(CC3, 19)
(CCSDT, 25)
(CCSDTQ-1a, 28)
(CCSDTQ-1b,  28)
(CCSDTQ-3,  22)
(CC4,  24)
(CCSDTQ, 28)};
\addplot coordinates {
(CCSDT-1a, 10)
(CCSDT-1b, 10)
(CCSDT-2, 10)
(CCSDT-3, 10)
(CC3, 19)
(CCSDT, 18)
(CCSDTQ-1a, 9)
(CCSDTQ-1b,  9)
(CCSDTQ-3,  10)
(CC4,  10)
(CCSDTQ, 18)};

\end{groupplot}

\draw([xshift=-0.6cm]plots4 c1r1.west) circle (0pt) node {(a)};
\draw([xshift=-0.6cm]plots4 c1r2.west) circle (0pt) node {(b)};
\draw([xshift=-0.6cm]plots4 c1r3.west) circle (0pt) node {(c)};
\draw([xshift=-0.6cm]plots4 c1r4.west) circle (0pt) node {(d)};

\end{tikzpicture}

\caption{\label{fig:all-lambda}Comparison of the number of iterations required
for convergence in the CC $\Lambda$ equations for full DIIS and the
suggested sub-iteration method (2 CCSDT sub-iterations and 3 CCSD
sub- or micro-iterations). Dynamic damping is included in the sub-iteration
CCSDT and CCSDTQ calculations. (a) $\ce{H2O}$, (b) $\ce{BeO}$, (c)
$\ce{O3}$, (d) $\ce{C2}$.}
\end{figure}
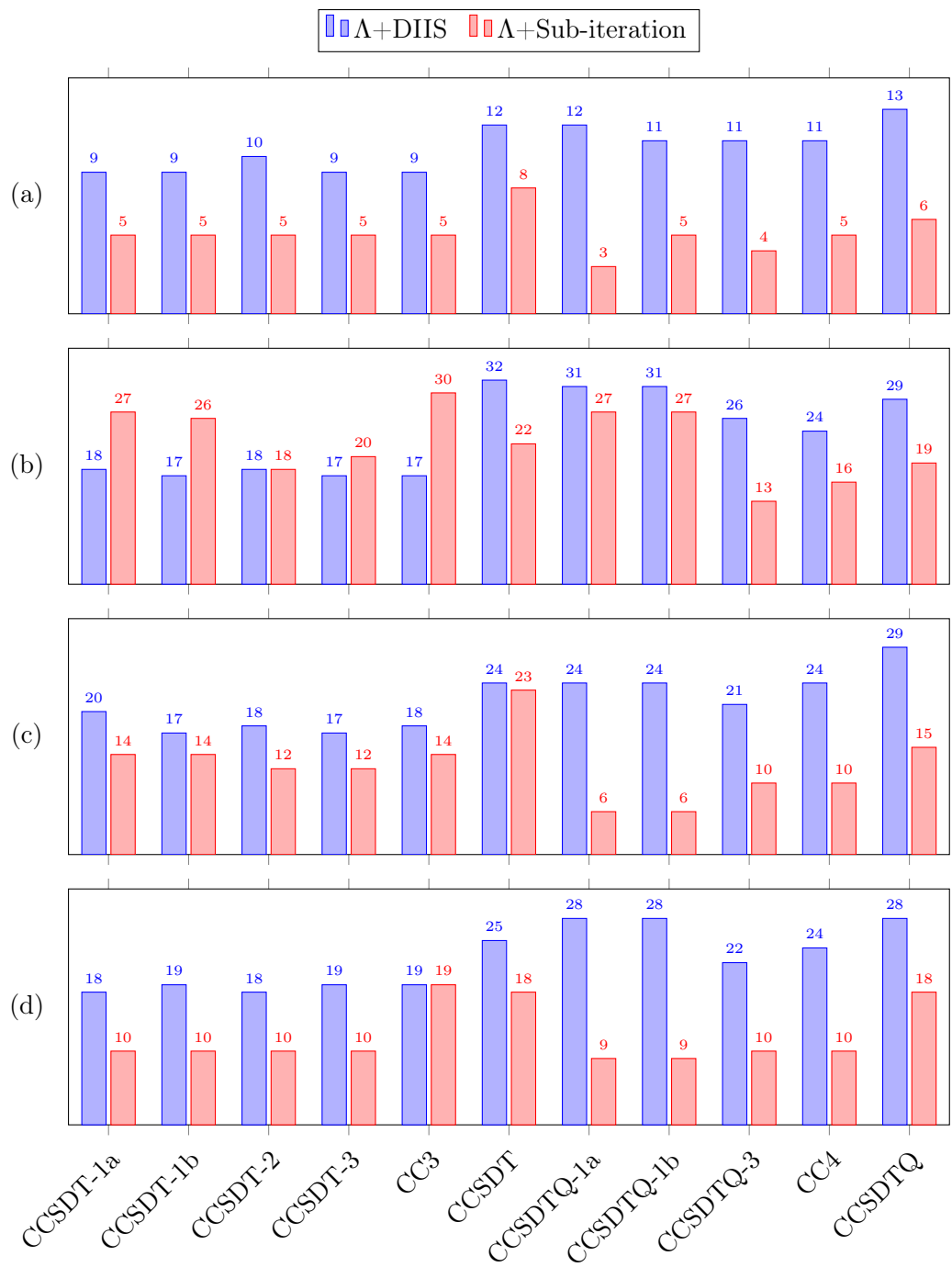

$\ce{H2O}$ is expected to provide a prototypical picture of the convergence
properties for well-behaved molecules, i.e. single-reference closed-shell
systems with weak dynamic correlation. Of course, such systems are
also the easiest for which to solve the coupled cluster and $\Lambda$
equations. In order to provide a more challenging test case, we also
examine the behavior of the sub-iteration methods for $\ce{BeO}$,
$\ce{O3}$, and $\ce{C2}$. BeO exhibits an extremely large $\hat{T}_{1}$
amplitude, approximately 0.13, as well as a large $\hat{\Lambda}_{1}$
amplitude, approximately 0.09. Both $\ce{O3}$ and $\ce{C2}$ exhibit
very large $\hat{T}_{2}$ amplitudes, approximately 0.32 and 0.38,
as well as large $\hat{\Lambda}_{2}$ amplitudes, 0.26 and 0.3, respectively.
Both symptoms are commonly attributed to multi-reference character,
of which the latter two are common examples. Such systems represent
very challenging tests for convergence of the coupled cluster equations,
as is seen in the data presented here. In Paper I, it was observed
that large $\hat{T}_{2}$ amplitudes did not seem to be an impediment
to the sub-iteration method, but that BeO, presumably due to the large
$\hat{T}_{1}$ amplitudes, lead to a severe degradation or complete
lack of convergence when using sub-iteration. This lead to the development
of the dynamic damping method proposed here. Figures \ref{fig:all-cc}
and \ref{fig:all-lambda} summarize the results for all four molecules
using the default settings described above. For the amplitude equations,
there are two notable differences with respect to the result presented
in Paper I. First, the subtle changes to the sub-iteration algorithm
described in \subsecref{Implementation-Details} lead to slightly
better performance of the approximate triples methods applied to $\ce{H2O}$.
Previously we measured little or no improvement for these methods,
but here we see a consistent decrease in the number of iterations
by about 40\%. Second, for BeO, we see a much improved situation due
to dynamic damping. While the approximate triples methods still struggle
(these methods do no benefit from dynamic damping), the CCSDT and
CCSDTQ results are significantly better than those in Paper I. For
$\ce{O3}$ and $\ce{C2}$, consistently improved convergence due to
sub-iteration is seen as before. The results for the $\Lambda$ equations
are remarkably similar, although the absolute number of iterations
required is rather smaller in all cases compared to the amplitude
equations. This effect could be attributed to the fact that the $\Lambda$
equations are strictly linear. One outlier in both figures is the
case of CCSDT for $\ce{O3}$. As will be seen below, this is due to
dynamic damping.

\begin{figure}
\begin{tikzpicture}
\begin{axis}[
ymin=0,
ymax=100,
ybar=2,
width=\textwidth,
height=6cm,
xtick=data,
enlarge x limits=.15,
bar width=10,
xticklabels={$\ce{H2O}$, $\ce{BeO}$, $\ce{O3}$, $\ce{C2}$},
legend columns=1,
ylabel={\% of Iterations Damped},
legend style={/tikz/every even column/.append style={column sep=0.25cm}},
legend style={legend pos=north east},
bar width=14,
x tick label style={rotate=45,anchor=north east}
]
\addplot coordinates {
(0, 0.1)
(1, 46.9)
(2, 0.1)
(3, 0.1)
};
\addplot coordinates {
(0, 0.1)
(1, 68.4)
(2, 25)
(3, 3.4)
};
\addplot coordinates {
(0, 0.1)
(1, 51.4)
(2, 92.2)
(3, 10.3)
};
\addplot coordinates {
(0, 0.1)
(1, 78.8)
(2, 20)
(3, 0.1)
};
\legend{CCSDT $\hat T$, CCSDTQ $\hat T$, CCSDT $\hat \Lambda$, CCSDTQ $\hat \Lambda$}

\end{axis}

\draw (8.3,4.15) circle (0pt) node {*};

\end{tikzpicture}

\caption{\label{fig:Fraction-of-iterations}Fraction of iterations in the CCSDT
or CCSDTQ amplitude and $\Lambda$ equations that required damping.
These calculations were converged to a tolerance of $10^{-10}$, but
the damping pattern across iterations is fairly consistent. The starred
calculation is discussed in the text.}
\end{figure}
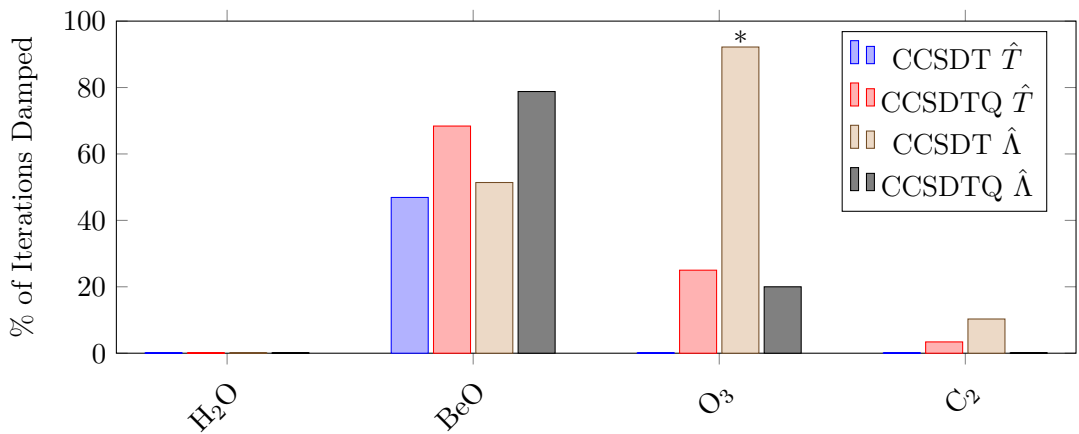

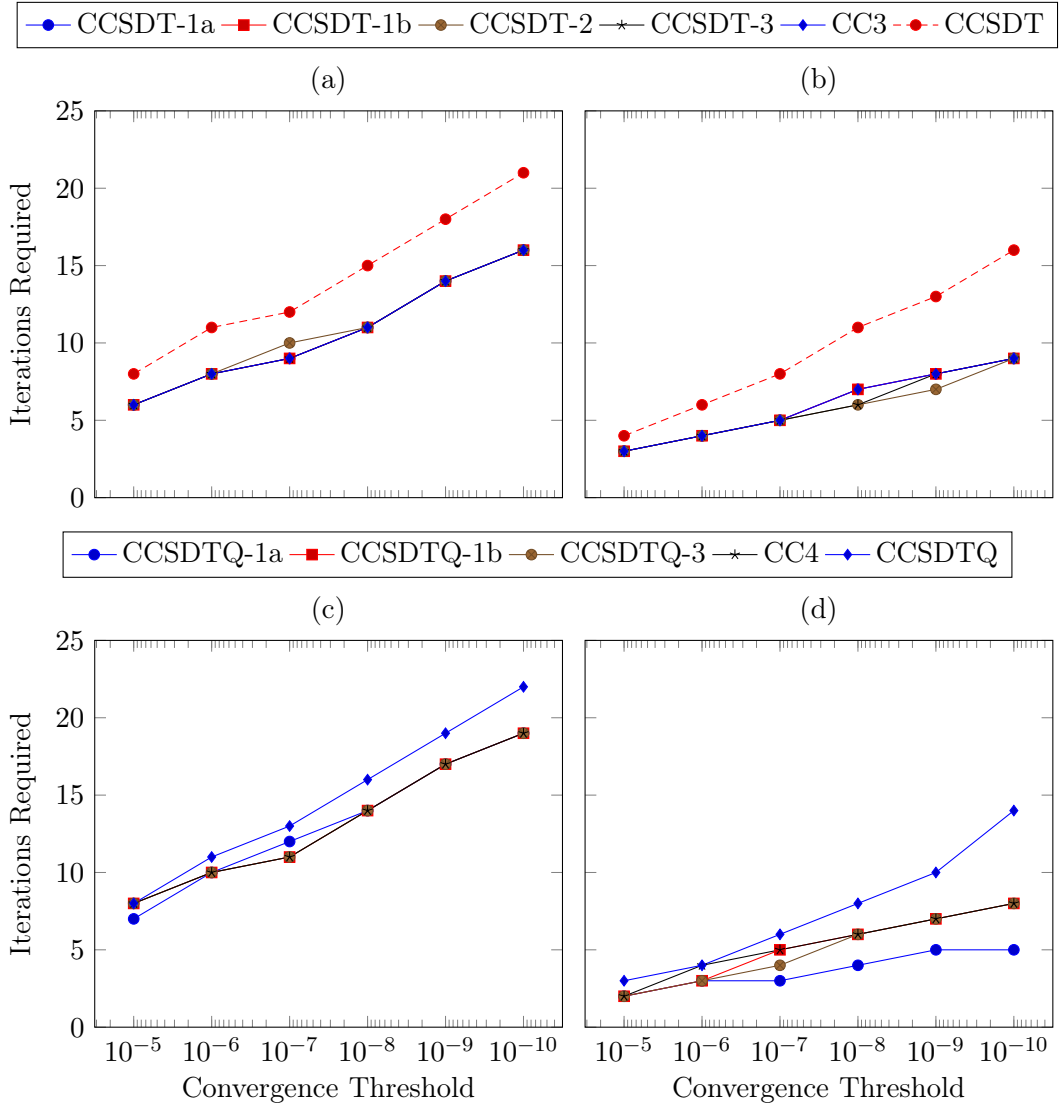
\begin{figure}
\begin{tikzpicture}
\pgfplotsset{try min ticks=7}
\begin{groupplot}[
width=7.8cm,
group style={y descriptions at=edge left,x descriptions at=edge bottom,group size=2 by 2,group name=plots,horizontal sep=0.3cm,vertical sep=1.9cm},
ymin=0,
ymax=25,
xmode=log,
x dir=reverse,
ylabel near ticks,
xlabel={Convergence Threshold},
ylabel={Iterations Required}
]

\nextgroupplot[
legend to name=legend1,
legend columns=-1]
\legend{CCSDT-1a, CCSDT-1b, CCSDT-2, CCSDT-3, CC3, CCSDT, CCSDTQ-1a, CCSDTQ-1b, CCSDTQ-3, CC4, CCSDTQ}
\addplot +[ sharp plot ] coordinates {
(1e-5, 6)
(1e-6, 8)
(1e-7, 9)
(1e-8, 11)
(1e-9, 14)
(1e-10, 16)};
\addplot +[ sharp plot ] coordinates {
(1e-5, 6)
(1e-6, 8)
(1e-7, 9)
(1e-8, 11)
(1e-9, 14)
(1e-10, 16)};
\addplot +[ sharp plot ] coordinates {
(1e-5, 6)
(1e-6, 8)
(1e-7, 10)
(1e-8, 11)
(1e-9, 14)
(1e-10, 16)};
\addplot +[ sharp plot ] coordinates {
(1e-5, 6)
(1e-6, 8)
(1e-7, 9)
(1e-8, 11)
(1e-9, 14)
(1e-10, 16)};
\addplot +[ sharp plot ] coordinates {
(1e-5, 6)
(1e-6, 8)
(1e-7, 9)
(1e-8, 11)
(1e-9, 14)
(1e-10, 16)};
\addplot +[ sharp plot ] coordinates {
(1e-5, 8)
(1e-6, 11)
(1e-7, 12)
(1e-8, 15)
(1e-9, 18)
(1e-10, 21)};

\nextgroupplot
\addplot +[ sharp plot ] coordinates {
(1e-5, 3)
(1e-6, 4)
(1e-7, 5)
(1e-8, 7)
(1e-9, 8)
(1e-10, 9)};
\addplot +[ sharp plot ] coordinates {
(1e-5, 3)
(1e-6, 4)
(1e-7, 5)
(1e-8, 7)
(1e-9, 8)
(1e-10, 9)};
\addplot +[ sharp plot ] coordinates {
(1e-5, 3)
(1e-6, 4)
(1e-7, 5)
(1e-8, 6)
(1e-9, 7)
(1e-10, 9)};
\addplot +[ sharp plot ] coordinates {
(1e-5, 3)
(1e-6, 4)
(1e-7, 5)
(1e-8, 6)
(1e-9, 8)
(1e-10, 9)};
\addplot +[ sharp plot ] coordinates {
(1e-5, 3)
(1e-6, 4)
(1e-7, 5)
(1e-8, 7)
(1e-9, 8)
(1e-10, 9)};
\addplot +[ sharp plot ] coordinates {
(1e-5, 4)
(1e-6, 6)
(1e-7, 8)
(1e-8, 11)
(1e-9, 13)
(1e-10, 16)};

\nextgroupplot[
legend to name=legend2,
legend columns=-1]
\legend{CCSDTQ-1a, CCSDTQ-1b, CCSDTQ-3, CC4, CCSDTQ}
\addplot +[ sharp plot ] coordinates {
(1e-5, 7)
(1e-6, 10)
(1e-7, 12)
(1e-8, 14)
(1e-9, 17)
(1e-10, 19)};
\addplot +[ sharp plot ] coordinates {
(1e-5, 8)
(1e-6, 10)
(1e-7, 11)
(1e-8, 14)
(1e-9, 17)
(1e-10, 19)};
\addplot +[ sharp plot ] coordinates {
(1e-5, 8)
(1e-6, 10)
(1e-7, 11)
(1e-8, 14)
(1e-9, 17)
(1e-10, 19)};
\addplot +[ sharp plot ] coordinates {
(1e-5, 8)
(1e-6, 10)
(1e-7, 11)
(1e-8, 14)
(1e-9, 17)
(1e-10, 19)};
\addplot +[ sharp plot ] coordinates {
(1e-5, 8)
(1e-6, 11)
(1e-7, 13)
(1e-8, 16)
(1e-9, 19)
(1e-10, 22)};

\nextgroupplot
\addplot +[ sharp plot ] coordinates {
(1e-5, 2)
(1e-6, 3)
(1e-7, 3)
(1e-8, 4)
(1e-9, 5)
(1e-10, 5)};
\addplot +[ sharp plot ] coordinates {
(1e-5, 2)
(1e-6, 3)
(1e-7, 5)
(1e-8, 6)
(1e-9, 7)
(1e-10, 8)};
\addplot +[ sharp plot ] coordinates {
(1e-5, 2)
(1e-6, 3)
(1e-7, 4)
(1e-8, 6)
(1e-9, 7)
(1e-10, 8)};
\addplot +[ sharp plot ] coordinates {
(1e-5, 2)
(1e-6, 4)
(1e-7, 5)
(1e-8, 6)
(1e-9, 7)
(1e-10, 8)};
\addplot +[ sharp plot ] coordinates {
(1e-5, 3)
(1e-6, 4)
(1e-7, 6)
(1e-8, 8)
(1e-9, 10)
(1e-10, 14)};

\end{groupplot}

\node at (plots c1r1.north east) [inner sep=0pt,anchor=south, yshift=5ex,xshift=-2ex] {\ref{legend1}};
\node at (plots c1r2.north east) [inner sep=0pt,anchor=south, yshift=5ex,xshift=-2ex] {\ref{legend2}};
\draw([yshift=0.4cm]plots c1r1.north) circle (0pt) node {(a)};
\draw([yshift=0.4cm]plots c2r1.north) circle (0pt) node {(b)};
\draw([yshift=0.4cm]plots c1r2.north) circle (0pt) node {(c)};
\draw([yshift=0.4cm]plots c2r2.north) circle (0pt) node {(d)};

\end{tikzpicture}\caption{\label{fig:tight}Number of iterations required to reach convergence
in the CC $\Lambda$ equations ($\ce{H2O}$) as a function of convergence
threshold. The upper panels (a,b) depict the results of triples methods
with full DIIS and sub-iteration, respectively. The bottom panels
(c,d) depict the results of the quadruples methods. The equivalent
plots for the amplitude equations are not shown, but they have identical
qualitative behavior.}
\end{figure}

\figref{Fraction-of-iterations} illustrates to what degree dynamic
damping is activated in each case (note that dynamic damping only
affects CCSDT and CCSDTQ). For $\ce{H2O}$, no damping is observed
as expected due to the already very good convergence achieved by sub-iteration.
$\ce{C2}$ similarly exhibits almost no damping. The amplitude and
$\Lambda$ equations for BeO are damped between 40 and 80\% of the
time, with a higher fraction of iterations damped for CCSDTQ than
CCSDT. The damping pattern observed in these calculations is very
regular: as soon as oscillation begins to set in (as early as the
fourth or fifth iteration), damping is activated with a damping parameter
averaging between 0.4 and 0.6. Once the oscillation is damped out,
the damping is naturally removed and the equations proceed for several
iterations before the cycle repeats. Despite the need for a rather
large amount of damping, the sub-iteration method still manages to
improve rather drastically on full DIIS. $\ce{O3}$ is an outlier
in that it only experience significant damping for the CCSDT $\Lambda$
equations (marked with an asterisk in \figref{Fraction-of-iterations}).
It is not clear why damping is triggered in this case, but based on
a comparison to full DIIS, it is clear that it does not have a positive
effect on convergence. While at a convergence threshold of $10^{-7}$,
the sub-iteration calculation still manages to finish in fewer iterations
than full DIIS, at a tighter threshold of $10^{-10}$, sub-iteration
requires 51 iterations while DIIS requires only 40. While this particular
case is not fully understood, it seems that dynamic damping may be
employed in the general case with overall positive effect.

Finally, we note that, especially in the context of the coupled cluster
$\Lambda$ equations, a convergence threshold of $10^{-7}$ may not
be sufficient, e.g. for accurate geometry optimizations. So, how does
sub-iteration perform at much tighter convergence thresholds? To address
this question, the number of iterations required to converge the $\Lambda$
equations with thresholds varying between $10^{-5}$ and $10^{-10}$
are presented in \figref{tight}. From these results, we see that
the significant benefit of sub-iteration over full DIIS is evident
even out to very tight convergence thresholds. While explicit results
are only shown for $\ce{H2O}$, the story is similar for the other
molecules tested, even BeO. The once exception is the CCSDT $\Lambda$
calculation of $\ce{O3}$, which as noted before, experiences spurious
damping.

\section{Conclusions}

We have extended the sub-iteration method for higher-order coupled
cluster methods presented in Paper I to the case of the coupled cluster
$\Lambda$ equations. Additionally, we have implemented a dynamic
damping scheme based on the method of Zerner and Hehenberger in order
to cure oscillatory behavior encountered in the case of systems with
large $\hat{T}_{1}$ amplitudes (and potentially other cases as well).
The good performance of sub-iteration for the amplitude equations
observed in Paper I is closed reflected in the results for the $\Lambda$
equations, and with the addition of dynamic damping, the sub-iteration
method is very nearly universally more effective in rapidly converging
the coupled cluster amplitude and $\Lambda$ equations compared to
full DIIS. The sub-iteration method is not conceptually complex, but
the specific implementation of the method required to obtain the best
results depends on several subtle points summarized in \subsecref{Implementation-Details}.
The availability of an effective and cheap method (in terms of memory
and disk I/O) that can ensure a rapid convergence of higher-order
coupled cluster methods, especially in the important cases of analytic
gradients and properties, is an important step in enabling the routine
use of such methods in computational chemistry.

\section*{Acknowledgments}

I would like to thank Prof. Jürgen Gauss for inspiring this work,
and for his efforts on the CCSDT implementation in CFOUR on which
this work is ultimately based. A preliminary version of the original
work on sub-iteration in the CC amplitude equations was also presented
at the 2015 workshop ``New Developments in Coupled-Cluster Theory''
co-organized by Prof. Gauss. This work was supported by a start-up
grant from Southern Methodist University, and all calculations were
performed on the ManeFrame II computing system at SMU.

\section*{Disclosure Statement}

No potential conflict of interest was reported by the authors.

\bibliographystyle{tfo}
\bibliography{convergence}

\end{document}